\documentclass[twocolumn,prb,showpacs,preprintnumbers,amsmath,amssymb]{revtex4}

\usepackage [dvips]{epsfig}
\usepackage {graphics}
\usepackage {graphicx}
\usepackage{graphicx}
\usepackage{dcolumn}
\usepackage{bm}

\begin{document}

\title{First-principles study of filled and unfilled skutterudite 
antimonies}

\author{Marek Veithen}
\author{Philippe Ghosez}
\affiliation{D\'epartement de Physique, Universit\'e de Li\`ege,
B-5, B-4000 Sart-Tilman, Belgium}

\date{\today}

\begin{abstract}

The electron localization tensor, Born effective charges, dielectric
constants and phonon dispersion relations
of the skutterudite CoSb$_3$
and the filled skutterudite TlFeCo$_3$Sb$_{12}$ 
have been studied using density functional perturbation theory. 
The origin of the low energy peak in the phonon density of states of
TlFeCo$_3$Sb$_{12}$ observed recently by Neutron inelastic
scattering 
[R. P. Hermann {\it et al.}, Phys. Rev. Lett. {\bf 90}, 135505 (2003)]
is attributed to the vibrations of Tl that
show only weak coupling with the normal modes of the host crystal.
Moreover, the dielectric properties of these materials show
unsual features such as giant Born effective charges and
a strong increase of the optical and static dielectric tensor
in the filled compound.

\end{abstract}



\maketitle

\section{Introduction}

CoSb$_3$ belongs to the class of binary skutterudites~\cite{scasm69_139}. 
These materials
are characterized by a complex cristalline structure containing
large voids and four formula units per cell. It has been observed
that filling the voids with electropositive atoms such as
Tl or rare earth atoms leads to a drastic decrease of the lattice
thermal conductivity~\cite{science272_1325,prb61_2475}. 
Combined with good semiconducting properties
(electrical conductivity, Seebeck coefficient), this makes
the so-called filled skutterudites potentially interesting for 
thermoelectric applications such as refrigeration or energy
generation~\cite{rppp51_459}.

One mechanism that has been evoked to explain
the strong reduction of the lattice thermal conductivity in filled
skutterudites is a strongly anharmonic rattling motion of the filling
atom. Keppens and co-workers~\cite{nature395_876} used inelastic
neutron scattering to measure the phonon density of states (DOS)
of the La atoms in LaFe$_4$Sb$_{12}$. They found a well defined peak
at about 56 cm$^{-1}$ and a somewhat broader structure around
121 cm$^{-1}$ that were associated to two La-dominated localized
modes. Feldman and co-workers~\cite{prb61_r9209}
used a force constant
model fitted from first-principles calculations
to compute the phonon dispersion relations and DOS
of LaFe$_4$Sb$_{12}$ and CeFe$_4$Sb$_{12}$. Their results allowed them to
refute the hypothesis of an anharmonic rare-earth potential and
to suggest that the two peaks in the La projected DOS are not
associated to two localized modes but are the results of hybridizations between
the rare-earth and Sb vibrations.

In several respects, we can
expect the behaviour of Tl to be different
from that of La and Ce.
First, the 4f valence states of the rare earth atoms
strongly hybridize with the transition metal d and Sb p states
in the vicinity of the Fermi energy~\cite{prb53_1103}.
These hybridizations are absent in case of Tl where the 4f
states are much lower in energy.
Second, the electronegativity of Tl is significantly
higher than that of La and Ce and close to that of Sb.
This suggests that Tl may have a smaller effect on the
electrical transport properties of CoSb$_3$ than the rare-earths.
Finally, the ionic radius of Tl is significantly
larger than that of La or Ce. 
We can therefore expect the dynamical properties of Tl to be
significantly different from that of the rare earths.

Some evidence of a different behaviour of Tl can be found in a
recent work of Hermann and co-workers~\cite{prl90_135505} who measured
the phonon DOS of various Tl-filled skutterudites. 
These authors found a single peak arount 40 cm$^{-1}$, 
in striking difference
with the two peaks of the La-filled compound.

Motivated by this work~\cite{prl90_135505},
we use first-principles calculations
to study the lattice dynamics and dielectric properties of
CoSb$_3$ and TlFeCo$_3$Sb$_{12}$. We show that Tl is in a
harmonic potential well up to large displacements.
Moreover, the peak about 40 cm$^{-1}$ in the phonon DOS of the
filled compound is due to the vibrational motion of Tl that
shows only modest hybridizations whith the normal modes of the
host crystal. Our calculations also confirm the existence
of a single Tl-dominated peak in the DOS of the filled compound.
In addition, we observe unusual dielectric porperties in CoSb$_3$
and TlFeCo$_3$Sb$_{12}$ such as giant Born effective charges,
or a strong increase of the static
dielectric constant in the filled compound.
To our knowledge, this is the first first-principles study
of Tl-filled skutterudites and the first time
Born effective charges and dielectric constants have been
computed from first-principles for these compounds.

Our paper is organized as follows. Sec. \ref{sec_tecdet} contains
the technical details of our calculations. In Sec. \ref{sec_gs},
we discuss the structural and electronic properties of CoSb$_3$
and TlFeCo$_3$Sb$_{12}$. In Sec. \ref{sec_dielprop}, we discuss
their optical dielectric constant and effective charges.
In Sec. \ref{sec_zonecenterphon}, 
we report the zone-center phonons in the filled and unfilled
compound and deduce their static dielectric tensor and infrared
reflectivity.
Finally, in Sec. \ref{sec_phon_disp}, we
discuss the phonon dispersion relations and DOS of 
CoSb$_3$ and TlFeCo$_3$Sb$_{12}$.


\section{Technical details} \label{sec_tecdet}

Our calculations have been performed within the local density approximation
to density functional theory (DFT)~\cite{hk,ks} thanks to the 
{\sc abinit}~\cite{abinit} package. 
We worked at the experimental lattice constant. The atomic positions have
been relaxed until the forces on the atoms were smaller than 
$5 \cdot 10^{-5}$ hartree/bohr.
For the exchange-correlation energy, we
used the parametrization of Perdew and Wang~\cite{prb45_13244}.
The valence states were computed non-relativistically.
The all-electron potentials were replaced by norm-conserving
pseudopotentials generated from scalar relativistic all-electron
calculations of the free atoms thanks to the FHI98PP code~\cite{fuchs}.
The Sb potential has been generated according to the Troullier-Martins
scheme~\cite{prb43_1993}. In order to obtain smooth pseudopotentials for
Fe, Co and Tl we used the Hamann~\cite{prb40_2980} scheme for these elements.
The parameters used to generate the pseudopotentials are summarized in
Table \ref{tab_psp}. 
The electronic wavefunctions have been expanded in a
plane wave basis. With the pseudopotentials described above
a kinetic energy cutoff of 25
hartree was enough to obtain well converged results. 

The static ionic charges of Sec. \ref{sec_zeffective}
have been computed following Bader's prescription
for the partitioning of space into atomic bassins~\cite{bader} as
it is described in Ref. \onlinecite{prb69_85411}.

For the phonon frequencies, a $2 \times 2 \times 2$ grid
of special k-points~\cite{prb13_5188} was sufficient to obtain
well converged results while the Born effective charges, dielectric
tensors and localization tensors required a grid of
$4 \times 4 \times 4$ special k-points. In fact, for the smaller grid,
the violation of the charge neutrality for the Born
effective charges~\cite{prb55_10355} 
was quite important (0.2 a. u.). Using the larger grid, this
error could be reduced by a factor of 10 while the phonon frequencies
changed by less than 3 cm$^{-1}$.

The phonon frequencies, Born effective charges, localization tensors
and dielectric tensors have
been computed from a linear response approach to 
DFT~\cite{prb55_10355,prb55_10337,prb66_235113}.
The first-order derivatives of the wavefunctions have been calculated
by minimizing a variational expression of the second-order
energy derivatives within the parallel gauge. In order to perform
the band-by-band decomposition of various quantities such as
the localization tensor or the optical dielectric tensor, these
wavefunctions were further transformed to the diagonal gauge~\cite{jpc12_9179}.
We computed the dynamical matrix explicitely on a $2 \times 2 \times 2$
grid of special q-points. To obtain the full phonon band structure and
DOS, we used
a Fourier interpolation that includes the long range dipole-dipole
interactions~\cite{prb50_13035}. 
The lattice specific heat has been computed from the DOS
as it is described in Ref. \onlinecite{prb51_8610}.
This approach is more reliable than the one
used in the previous studies of Singh and 
co-workers~\cite{prb53_6273,prb61_r9209,prb68_94301}.
In Refs. \onlinecite{prb53_6273} and \onlinecite{prb61_r9209}, 
these authors used 
an empirical force constant model that had been adjusted
throught LDA total energy calculations while in Ref. 
\onlinecite{prb68_94301}, they computed the interatomic force constants
for zone-center atomic displacements only.

\begin{table*}[htbp]
\caption{\label{tab_psp} Parameters used to generate the pseudopotentials
of Fe, Co, Sb and Tl. Reported is the type of the pseudopotential (H =
Hamann, TM = Troullier-Martins), the projector that is used as local
component, the cutoff radii (bohr) of the s, p, d and f channels (in case they
have been included in the pseudopotential) and 
the cutoff radius (bohr) of the partial core density (r$_{nlc}$).
The values in brackets indicate the reference energy (eV) that has been
used to build the pseudo-wavefunction in case we did not use 
the default value of the FHI98PP code.}
\begin{ruledtabular}
\begin{tabular} {lcccc}
               &      Fe       &       Co       &     Sb     &    Tl     \\
\hline
Type           &      H        &       H        &     TM     &    H      \\
Configuration  & 4s$^1$ 3d$^6$ 4p$^0$ & 4s$^1$ 3d$^7$ 4p$^0$ &
                 5s$^2$ 5p$^3$        & 6s$^2$ 5d$^{10}$ 6p$^0$ \\
Local part     &      f        &       f        &      s     &    p      \\
s-channel & 1.2 & 1.2            & 1.8        & 1.4 \\
p-channel & 1.2 (5.0)& 1.2 (5.0)      & 2.0        & 1.4 \\
d-channel & 1.0 & 1.0            & 2.5 (15.0) & 1.4 \\
f-channel & 1.6\footnotemark[1] & 
                 1.6\footnotemark[1] &
                 --                   & 
                 1.6\footnotemark[1] \\
r$_{nlc}$ & 0.9& 0.9         & 1.5        & 1.0 \\
\end{tabular}
\end{ruledtabular}
\footnotetext[1]{Default value used by the FHI98PP code}
\end{table*}


\section{Ground-state properties} \label{sec_gs}

\subsection{Crystal structure} \label{sec_struct}

The structure of CoSb$_3$ is formed by a bcc lattice 
[space group $Im\overline{3}$, lattice constant a =
9.0385 \AA] with 4 formula units per primitive unit cell.
It can be described as a distorted 
perovskite~\cite{scasm69_139}
with the chemical formula ABO$_3$. In CoSb$_3$, the site of
the A atoms is empty, the sites of the B atoms are occupied by Co and the
positions of O by Sb. The CoSb$_6$ octaedra are tilted so that the Sb
atoms form rectangular Sb$_4$ rings.
The Co and Sb atoms occupy respectively the 8c 
$(\frac{1}{4},\frac{1}{4},\frac{1}{4})$
and 24g 
$(0,y,z)$ Wyckoff positions. 
The relaxed values of $y = 0.33284$ and $z = 0.15965$ are in exellent agreement
with the experimental results~\cite{actacc43_1678} 
of $y = 0.33537$ and $z = 0.15788$.

In the unfilled compound, the 2a $(0,0,0)$ Wyckoff positions are empty. 
They are at the centers of large voids 
surrounded by 12 Sb and 8 Co atoms.
These voids are able to accomodate an
electropositive filler such as Tl or various rare earth atoms. 
The electronic configuration of Tl is 6s$^2$ 6p$^1$. In order to fill an
important fraction $x$ of voids, all chemical bonds must be saturated.
Therefore, one has to compensate the excess charge
due to the Tl 6p electrons by replacing for example one Co per
cell by Fe or one Sb by Sn~\cite{prb61_2475}.

Experimentally, the lattice constant of Tl-filled CoSb$_3$ is found to
increase linearly with the amount of filled voids~\cite{prb61_2475} for 
$0 < x < 0.8$. 
In our calculation, we worked at $x = 1$ and we replaced one Co per 
cell by Fe.
We chose a lattice
constant of 9.1276~\AA {\ }
that corresponds to a linear extrapolation of the data of
Ref. \onlinecite{prb61_2475} to $x = 1$.
This choice is further justified by the fact that the residual stress in
the filled compound ($2.7 \cdot 10^{-4}$ Ha/bohr$^3$) is similar
to the stress in the unfilled compound ($2.5 \cdot 10^{-4}$ Ha/bohr$^3$).
The relaxed reduced atomic coordinates in TlFeCo$_3$Sb$_{12}$
and CoSb$_3$
differ by less than 0.003.

Due to the use of periodic boundary conditions, the Fe atoms occupy the
same crystallographic site in each cell. This situation is somewhat
artificial since it neglects any disorder introduced by the 
random substitution of
Co by Fe in real samples. 
However, we expect the lack of disorder in our calculations
to have only modest
effects on the electronic and dynamical properties of Tl for the 
Fe-compensated compound. It might be more important for 
other compensations such as the substituaion of Sb by Sn.
First, Fe can only occupy four equivalent sites per cell
while there are 12 sites that can accomodate Sn. Second, the distance
between Tl and Sb (3.3897 and 3.3977 \AA) is smaller than the distance
between Tl and Co (3.9823 \AA). 
Finally, we will see in Sec. \ref{sec_phon_disp} 
that the vibrations of Co/Fe
are strongly decoupled from the vibrations of Tl
and located in the high energy region 
above 200 cm$^{-1}$ of the phonon spectrum.
At the opposite, the vibrations of Sb occupy the region
below 200 cm$^{-1}$ and show significant interactions
with the vibrations of Tl.

\subsection{Electronic properties} \label{sec_elec}

\begin{figure}[htb]
\begin{center}
\includegraphics[width=7cm]{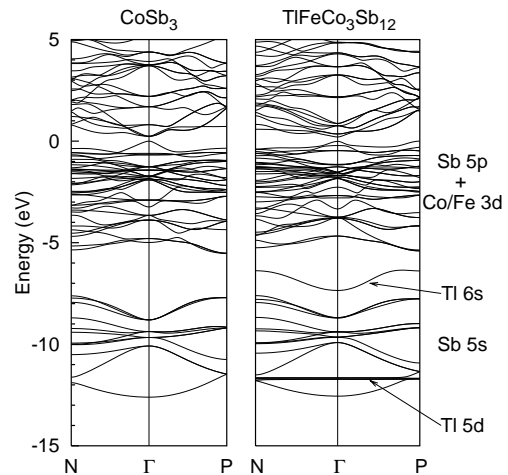}
\end{center}
\caption{\label{fig_electron} Electronic band structure in CoSb$_3$ and
TlFeCo$_3$Sb$_{12}$.}
\end{figure}

The left part of
Fig. \ref{fig_electron} shows the LDA band structure of CoSb$_3$ between
the high symmetry points N, $\Gamma$, and P. 
The top of the valence bands has been fixed at 0 eV.
There is a good agreement between our results and the band structures
obtained in previous 
studies~\cite{prb58_15620,prb50_11235,prb63_125110,jpc16_979}.
We observe the presence of two
well separated groups of valence bands. The bands between -13 eV and -7 eV are
mainly composed of Sb 5s states while the valence bands above -6 eV 
result from hybridizations between Sb 5p and Co 3d bands.

There is some controversy about the energy of the band gap, E$_g$,
in CoSb$_3$. Experimentally, values between 0.03 eV and 0.7 eV
have been reported
[see Refs. \onlinecite{jap80_4442,prb61_4672,apl76_3436} 
and references therein].
From a theoretical point of view, there is a similar strong
discrepancy between the values in the litterature.
The bandgap is defined by a highly dispersive band at the $\Gamma$-point.
We obtain a value of 0.24 eV in reasonable agreement with other
first-principles LDA and GGA calculations [0.05 eV~\cite{prb50_11235},
0.22 eV~\cite{prb58_15620}, 0.195 \& 0.330 eV~\cite{prb63_125110},
0.140 eV~\cite{jpc16_979}].
This discrepancy between the theoretical results has 
been attributed to a strong dependence of E$_g$ on the
position of the Sb atoms together with the well-known
problem of the LDA and GGA to predict the correct bandgap of 
semiconductors~\cite{prb58_15620,prb63_125110}.

In the TlFeCo$_3$Sb$_{12}$, [right part of Fig. \ref{fig_electron}], 
the bandgap (0.20 eV) is slightly smaller than in the unfilled compound.
In addition, we observe two additional groups of bands that are related
to the atomic orbitals of Tl. The Tl 5d states form a
flat set of bands that are in the same energy region as the Sb 5s states.
Due to the lack of dispersion, these states can be considered as 
inerte. At the opposite, the Tl 6s states form a band about -7 eV that shows a
significant dispersion related to a covalent interaction with the electronic
states of the host crystal.
This point will be investigated in more detail below.

In some cases, the filling of the voids in skutterudites does not only
affect the lattice thermal conductivity but also the electronic transport
properties. For example, the Ce 4f states in CeFe$_4$Sb$_{12}$ hybridize
with the Fe 3d and Sb 5p states to form a set of flat conduction bands in the
vincinity of the Fermi level~\cite{prb53_1103}. Due to the low dispersion
of these bands, the electron effective mass in this compound is quite high
implying a low electron mobility~\cite{prb56_15081}. 
As it can be seen in the right part of
Fig. \ref{fig_electron}, the effect of Tl filling on the electronic band
structure of CoSb$_3$ is less important. 
The only additional bands ly well below the Fermi level and will therefore
not affect the electronic transport properties. 
However, in order to conclude definitively about this point, 
it might be important 
to perform a more intensive study of the electronic band
structure of TlFeCo$_3$Sb$_{12}$ that eventually takes into account
spin-orbit coupling.

In order to get additional informations
on the electronic properties of CoSb$_3$ and TlFeCo$_3$Sb$_{12}$, 
we computed the localization tensor
(see Ref. \onlinecite{jpc14_r625} and references therein)
and the electronic density for selected states in these compounds.
As we have shown in a previous work~\cite{prb66_235113}, 
the decomposition of the localization tensor 
into contributions of individual groups of bands can act as a sensitive probe
to study hybridizations within a solid. For example, a state that has the
same variance in a solid and in the isolated atom can be considered as
chemically inerte. At the opposite, a modification of the variance is
usually attributed to covalent interactions of the corresponding
orbitals.

In CoSb$_3$ and TlFeCo$_3$Sb$_{12}$ the localization tensor is isotropic.
In Table \ref{tab_loctens}, we report the variances of the two
(resp. three) groups of bands defined in Sec. \ref{sec_elec}
as well as the variances of the atomic orbitals Sb 5s, Tl 5d and Tl 6s.
In case of Tl, we used linear response calculations to obtain the variance
of the 6s state of an isolated Tl$^+$ ion.
We adopted this configuration because it is close to the configuration we
expect for Tl in TlFeCo$_3$Sb$_{12}$. Moreover, it allowed us to avoid some
problems related to the linear response calculation of the localization
tensor for partially occupied states.
To use the same approach for Sb, we had to chose Sb$^{3+}$ as reference
configuration. For the 5s state, we obtained a value of 1.51 bohr$^2$.
Unfortunately, the character of the bonding 
of Sb in CoSb$_3$ is rather covalent than ionic.
The Sb$^{3+}$ configuration 
seems therefore somewhat artificial. To compute the variance for the
neutral Sb atom, we used the
FHI98PP~\cite{fuchs} code to compute the
all-electron wavefunction of the 5s state
and its variance as a matrix element of the squared position 
operator $x^2$.

\begin{table}[htbp]
\caption{\label{tab_loctens} Variances 
$\langle r^2 \rangle_c$
(Bohr$^2$) of the 
groups of bands in CoSb$_3$ and TlFeCo$_3$Sb$_{12}$ and of the Sb 5s,
Tl 5d and Tl 6s electrons
calculated on an isolated Sb atom and Tl$^+$ ion.}
\begin{ruledtabular}
\begin{tabular} {lcccccc}
Bands         & & CoSb$_3$ & & TlFeCo$_3$Sb$_{12}$ & & Atom \\
\hline
Sb 5s (+ Tl 5d) & & 3.06     & &  2.63               & & 1.86\footnotemark[1], 
                                                         0.85\footnotemark[2] \\
Tl 6s           & & ---      & & 18.06               & & 2.14   \\
Sb 5p + Co 3d   & & 5.11     & &  5.85               & & ---    \\
\hline
E$_g$ (eV)      & & 0.24     & &  0.20               & & ---    \\
\end{tabular}
\end{ruledtabular}
\footnotetext[1]{Sb 5s orbital of an isolated Sb atom.}
\footnotetext[2]{Tl 5d orbitals of an isolated Tl$^+$ ion.}
\end{table}

In CoSb$_3$, the variance of the Sb 5s bands is larger than the variance of
the corresponding atomic orbital. 
This result confirms the participation of these
electrons in the Sb-Sb bonding observed previously~\cite{prb63_125110}.
In the filled compound, the Tl 5d states occupy the same energy region as
the Sb 5s states. Therefore, the formalism of 
Ref. \onlinecite{prb66_235113} cannot be applied
to study the localization of these bands individually.
Nevertheless, we can obtain some information on the modification of the Sb
5s states in TlFeCo$_3$Sb$_{12}$. 
Due to the flat dispersion of the Tl 5d bands observed in Fig. 
\ref{fig_electron}, it is reasonable to assume that their variance in 
the solid is close to the value in the isolated Tl$^{+}$ ion.
If we neglect the covariance between Sb 5s and Tl 5d bands,
we obtain a value of 3.37 Bohr$^2$ for the variance of the Sb 5s
bands in TlFeCo$_3$Sb$_{12}$. This result suggests that the
Sb 5s electrons are more delocallized in the filled compound than in the
unfilled. 
It remains true in case the covariance between Sb 5s 
and Tl 5d bands is not zero since this quantity is necessarily
negative.

A stronger variation is found for the variance of the highest occupied
bands (Sb 5p + Co 3d). In TlFeCo$_3$Sb$_{12}$, these electrons are more
delocallized than in CoSb$_3$.

\begin{figure}[htb]
\begin{center}
\includegraphics[width=8cm]{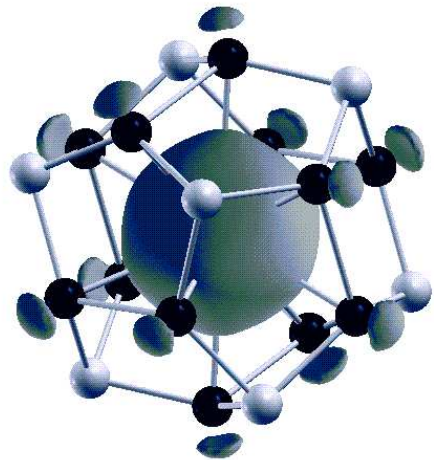}
\end{center}
\caption{\label{fig_density} Electronic pseudo-density of the Tl 6s band in
TlFeCo$_3$Sb$_{12}$. The isodensity value corresponds to 10 \% of the maximum
density of this state (0.0013 electron/bohr$^3$). Tl (not shown) occupies
the center of the cage. The black and grey spheres represent respectively
Sb and Co/Fe. 
This figure has been
realized thanks to the XCrySDen~\cite{xcrysden} crystalline and molecular
structure visualisation program.}
\end{figure}

The most spectacular evolution undergoes the variance of the Tl 6s state.
In the solid, this quantity is about 9 times larger than in the isolated
ion. This result suggests a significant interaction between the Tl 6s 
orbitals and the electronic states of the host crystal. 
To get more informations about these hybridizations we show
in Figure \ref{fig_density}
the electronic pseudo-density associated to the Tl 6s band.
As can be seen, these electrons are partially delocallized on the Sb
atoms of the cage. This result suggests that there is a significant
covalent interaction between Tl and Sb due to the Tl 6s electrons.
Such an interaction would be coherent with the fact that Tl can 
be inserted into the voids of the host crystal in
spite of the fact that its ionic radius (1.84 \AA) is larger
than that of the rare-earth fillers Ce (1.28 \AA) and La (1.50 \AA) 
and that it is close to
the radius of the cage (about 1.892 \AA~\cite{jap79_4002}).

In order to obtain further insight on the electronic
properties of skutterudites and the effect of different filling atoms
on the band structure
it might be interesting to build generalized Wannier 
functions~\cite{prb56_12847,prb65_35109} for these compounds.

In summary, the main differences between the electronic band structures
of CoSb$_3$ and TlFeCo$_3$Sb$_{12}$
affect the region well below the Fermi level. On the one hand, the Tl 5d
states show a flat dispersion suggesting that these states are chemically
inerte. On the other hand, the Tl 6s states show a significant dispersion
related to a partial delocalization of the corresponding electrons
on the Sb atoms of the cage. These hybridizations are expected to
affect the dielectric and dynamic properties of Tl
in the filled compound.


\section{Dielectric properties} \label{sec_dielprop}

\subsection{Optical dielectric tensor} \label{sec_eldiel}

In CoSb$_3$, the dielectric tensor is isotropic.
The theoretical value 
$\varepsilon_{\infty} = 31.67$ overestimates
the experimental result~\cite{irp24_171} of 25.6 by about 20 \%
as it is typical
in LDA calculations~\cite{prl74_4035}.
Similar large dielectric constants as in CoSb$_3$
have been computed for other narrow bandgap 
semiconductors such as Al$_2$Ru [principal values:
18.9, 22.9, 20.7]~\cite{prb54_r8297}. 
Moreover, they have been measured in a series of skutterudites such as
CoAs$_3$ (26.25)~\cite{jpcs49_267}, UFe$_4$P$_{12}$ (17) or 
CeFe$_4$P$_{12}$ (31)~\cite{prb60_11321}.

In the filled compound, the dielectric tensor is no more isotropic
but has weak off-diagonal elements. These elements are
about four orders of magnitude smaller than the diagonal ones
and we will not consider them in the discussion that follows.
We obtain a value of 38.57 for the dielectric
constant in TlFeCo$_3$Sb$_{12}$ that is significantly larger than 
in unfilled CoSb$_3$.
Unfortunatley, we are not aware of any experimental data
to which we can compare this result.
In spite of the error on its absolute value,
we expect the evolution of the dielectric constant 
to be correctly reproduced by our calculation
since it is reasonable to assume that the LDA error 
on $\varepsilon_{\infty}$ is
of the same magnitude in both compounds.

The increase of $\varepsilon_{\infty}$ can have distinct origins.
On the one hand, based on a simple Clausius-Mossotti model~\cite{kittel}, 
the polarizabilities of the Tl$^+$ ions
add to the dielectric susceptibility of the unfilled
compound to yield a larger $\varepsilon_{\infty}$ 
in TlFeCo$_3$Sb$_{12}$. On the other hand, the increase can be due
to the dielectric constant of the host crystal that is increased
due to the presence of Tl and Fe in the filled compound.
To check which mechanism applies in CoSb$_3$, we 
decomposed the optical dielectric tensor into contributions
originating from the individual groups of bands of Fig. \ref{fig_electron}.
This decomposition reveals that the deeper bands (Sb 5s, Tl 5d and Tl 6s)
are only weakly polarizable by an electric field and contribute
less than than 0.1 \% to $\varepsilon_{\infty}$
and that the contribution of the Sb 5p and transition metal 3d electrons
is strongly increased in the filled compound.
We therefore propose that the observed increase of $\varepsilon_{\infty}$
is due to a modification of the dielectric constant
of the host crystal while the
polarizability of Tl$^+$ plays only a minor role.

\subsection{Effective ionic charges} \label{sec_zeffective}

In Table \ref{tab_zbader}, we report the static ionic charges
Q$_i$ in CoSb$_3$ and TlFeCo$_3$Sb$_{12}$ obtained from a topological
analysis following Bader's prescription for the partitioning
of space into atomic bassins~\cite{bader}. 
In this approach, one has to determine surfaces that obey the zero-flux
condition for the electron density $n(\textbf{r})$:
$\bm{\nabla}n \cdot \bm{N} = 0$ where $\bm{N}$ is the vector normal to
the surfaces. The integration of $n(\textbf{r})$ within the atomic 
bassins yields the ionic charge Q$_i$ of each topological atom $i$.

In both compounds the chemical bonds have a negligible
ionic character as it is revealed by the small Q$_i$. 
At the opposite, to these small static charges, the dynamical 
Born effective charges $Z^{\ast}$ (Table \ref{tab_zast})
are very large, comparable to the giant effective charges
in ferroelectric ABO$_3$ compounds ~\cite{prl72_3618} and other narrow bandgap
semiconductors such as Al$_2$Ru~\cite{prb54_r8297}.
In the ferroelectrics, the amplitude of the $Z^{\ast}$
can be explained thanks to the Harrison model~\cite{harrison}.
A relative displacement of the B and O atoms
generates a giant dipole moment due to a dynamical 
change of hybridization between O 2p
and B d atomic orbitals~\cite{prb58_6224}.
In CoSb$_3$, the highest valence bands are basically due to hybridizations
between Co 3d and Sb 5p atomic orbitals~\cite{prb63_125110}.
Moreover, the structure of this compound
is formed of distorted CoSb$_6$ octahedra similar
to the BO$_6$ octahedra of the ABO$_3$ ferroelectrics.
It seems therefore plausible to assume that a similar mechnisme
takes place in CoSb$_3$
and that the large effective charges are generated by dynamic changes
of hybridizations between Co 3d and Sb 5p atomic orbitals.

In skutterudites, giant effective charges have been measured
from infrared spectroscopie
for UFe$_4$P$_{12}$ 
($|Z^{\ast}_{Fe}| = 8.9$) 
and CeFe$_4$P$_{12}$ ($|Z^{\ast}_{Fe}| = 11$)~\cite{prb60_11321}.
In CoSb$_3$, we obtain~\cite{notediel} a value of 6.04 for $|Z^{\ast}_{Co}|$
from the amplitude of the LO-TO splitting reported in 
Ref.~\onlinecite{pssb112_549} and the optical dielectric constant of
Ref.~\onlinecite{irp24_171}. This value is in good agreement with the results
reported in Tab. \ref{tab_zast}.

\begin{table}[htbp]
\caption{\label{tab_zbader} Bader ionic charges Q$_i$ ($|e^-|$)
in CoSb$_3$ and TlFeCo$_3$Sb$_{12}$.}
\begin{ruledtabular}
\begin{tabular} {lrclr}
       & CoSb$_3$ & &        & TlFeCo$_3$Sb$_{12}$ \\
\hline
       &          & & Tl     &  0.29                \\
Co$_1$ & -0.53    & & Fe     & -0.36                \\
Co$_2$ & -0.53    & & Co$_2$ & -0.48                \\
Sb     &  0.17    & & Sb     &  0.12                \\
\end{tabular}
\end{ruledtabular}
\end{table}

\begin{table}[htbp]
\caption{\label{tab_zast} Born effective charges ($|e^-|$)
in CoSb$_3$ and 
TlFeCo$_3$Sb$_{12}$.}
\begin{ruledtabular}
\begin{tabular} {lrrrclrrr}
& \multicolumn{3}{c}{CoSb$_3$} & & & 
\multicolumn{3}{c}{TlFeCo$_3$Sb$_{12}$} \\
\hline
       &      &      &      & &Tl     &  3.64 & -0.05 & -0.02 \\
       &      &      &      & &       & -0.02 &  3.64 & -0.05 \\
       &      &      &      & &       & -0.05 & -0.02 &  3.64 \\
       &      &      &      & &       &       &       &       \\
Co$_1$ &-6.89 & 0.82 & 0.34 & &Fe     & -9.83 &  1.13 & -0.20 \\
       & 0.34 &-6.89 & 0.82 & &       & -0.20 & -9.83 &  1.13 \\
       & 0.82 & 0.34 &-6.89 & &       &  1.13 & -0.20 & -9.83 \\
       &      &      &      & &       &       &       &       \\
Co$_2$ &-6.89 &-0.82 &-0.34 & &Co$_2$ & -7.28 & -0.75 &  0.08 \\
       &-0.34 &-6.89 & 0.82 & &       & -0.09 & -7.30 &  0.73 \\
       &-0.82 & 0.34 &-6.89 & &       & -0.64 & -0.10 & -7.34 \\
       &      &      &      & &       &       &       &       \\
Sb     & 2.63 & 0.00 & 0.42 & &Sb     &  2.43 & -0.10 & -0.04 \\
       & 0.00 & 1.94 & 0.00 & &       & -0.03 &  1.87 &  0.01 \\
       & 0.68 & 0.00 & 2.32 & &       &  1.01 &  0.12 &  2.13 \\
\end{tabular}
\end{ruledtabular}
\end{table}

The effective
charge of the filling Tl atom is also highly anomalous. It is significantly 
larger than its static ionic charge reported in Table 
\ref{tab_zbader}. 
Moreover, it is larger than its nominal ionic charge of +1
that would be expected if there was
a complete transfer of the 6p electron to the atoms
of the cage.
Giant effective charges have been reported previously~\cite{prb60_11321}
for the filling atoms U and Ce in FeP$_4$. Their
values of $Z^{\ast}_{U}=8.9$ and $Z^{\ast}_{Ce}=8.7$ are even larger 
than the value of $Z^{\ast}_{Tl}$ in CoSb$_3$.

The effective charges of Co and Sb are similar in the filled and the
unfilled compound. The excess effective charge introduced by Tl 
can therefore not be compensated by these
two atoms in order to satisfy the charge neutrality 
condition~\cite{prb55_10355}
$\sum_{\kappa} Z^{\ast}_{\kappa, \alpha \beta} = 0$.
As can be seen in Table \ref{tab_zast}, $|Z^{\ast}_{Fe}|$
is about 
$3 |e^-|$
larger than the 
absolute value of the effective charge
of the Co atom that it substitutes. As mentioned in Sec. \ref{sec_struct},
the 6p electron of Tl is compensated by a hole in the 3d orbitals of Fe.
In case of the Born effective charges, Fe plays a similar role:
due to the fact that $|Z^{\ast}_{Fe}|$ is larger than $|Z^{\ast}_{Co}|$,
the anomalous effective charge of Tl is mainly compensated by 
the anomalous effective charge of Fe.
A plausible mechanism for this compensation could be that
the Tl 6p electron is partially shared between Tl and Fe
so that it might be responsible for a transfer of charge
during a relative displacement of these two atoms.


\section{Zone-center phonons} \label{sec_zonecenterphon}

\subsection{Potential energy} \label{sec_potential}

To understand the effect of Tl-filling on the lattice thermal conductivity
of CoSb$_3$, it is mandatory to investigate the dynamics of these atoms in
their cages. Experimentally, unusually large atomic displacement parameters
(ADP's) have been observed for various filling 
atoms~\cite{prb56_15081,prb61_2475}.
To decide whether these ADP's are due to quasi-harmonic vibrations about the
center of the cage or to the random hopping between off-center 
sites~\cite{jap79_4002}, it is
important to investigate the shape of the potential energy well.

\begin{figure}[htb]
\begin{center}
\includegraphics[width=7cm]{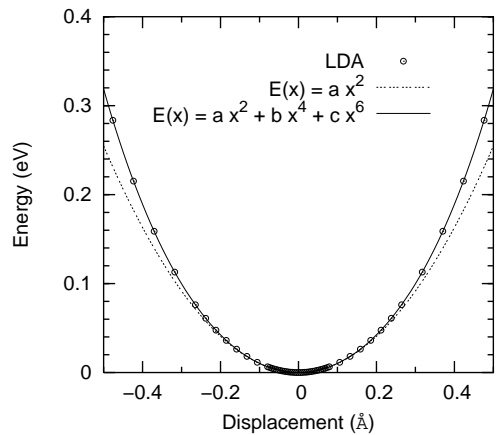}
\end{center}
\caption{\label{fig_pot} Potential energy associated to a cooperative
displacement of Tl along [1 0 0].}
\end{figure}

To determine the shape of the Tl potential, we computed the variations 
of the total energy for atomic displacements along various directions.
Figure \ref{fig_pot} shows the results for the cristallographic
direction [1 0 0].
We see that the center of the cage is the global minimum of the potential
energy well and not a saddle point 
as it has been suggested previously~\cite{jap79_4002}.
Close to the origin, the energy changes quadratically with the atomic
positons [dashed line].
The whole energy region of Fig. \ref{fig_pot}
can be fit accurately by a polynome of degree six [full line].
The harmonic frequency is 52 cm$^{-1}$. This so-called "bare frequency" is
quite independent of the direction of the atomic displacement. For the
directions [1 1 0] and [1 1 1], we obtaiend values that differ by less than
0.3 cm$^{-1}$ from the one of Fig. \ref{fig_pot}.
In case of Tl, the bare frequency is smaller than the one of Ce (68 cm$^{-1}$)
and La (74 cm$^{-1}$) in FeSb$_4$~\cite{prb61_r9209}.
This difference is attributed to the larger mass of Tl and to the
curvature of the potential energy at the origin that is significantly
smaller in case of Tl (2.0371 eV/\AA)
than for Ce
(2.3943 eV/\AA) and La (2.8005 eV/\AA).
This result is surprising. On the basis of the crystal 
radii~\cite{actaca32_751} of Tl$^{1+}$ (1.84 \AA), Ce$^{4+}$ (1.28 \AA) and
La$^{3+}$ (1.50 \AA) in a 12-coordinate site we could expect the repulsion
betwenn ionic cores to be more important between Tl and Sb than between
Ce/La and Sb. The eneregy should therefore increase more rapidly for an
off-center displacement of Tl and the curvature of the
potential energy well should be stronger.
However, in Sec. \ref{sec_elec} we saw that there is a significant
covalent interaction between Tl and Sb. The simple argument of the crystal
radii can therefore not be applied in case of TlFeCo$_3$Sb$_{12}$
because the atoms do not behave like rigid spheres.

The large ADP's of Tl have been attributed to an Einstein 
oscillator~\cite{prb61_2475} with a
characteristic frequency of 36 cm$^{-1}$.
This value is close to the Einstein frequencies deduced from specific heat
measurements and inelastic neutron scattering
experiments~\cite{prl90_135505} (39 $\pm$ 1 cm$^{-1}$)
but it is significantly lower than the computed Tl bare
frequency. This difference between the bare frequency and the
normal mode frequency observed experimentally will be 
further discussed in Sec. \ref{sec_phon_freq}.

\subsection{Normal mode frequencies} \label{sec_phon_freq}

The unfilled skutterudite CoSb$_3$ belongs to the space group $Im\overline{3}$.
At the $\Gamma$-point, the zone center optical phonons can be classified
according to its irreductible representations into
\begin{equation} \label{eq_phon}
2 A_g + 2 E_g + 4 F_g + 2 A_u + 2 E_u + 7 F_u.
\end{equation}
Only the F$_u$-modes are infrared active. At the $\Gamma$-point,
they are split into transverse (TO) and longitudinal (LO) modes. 
The $A_g$, $E_g$ and $F_g$ are Raman active.

In Table \ref{tab_phon}, we report the frequencies (cm$^{-1}$)
of the TO and LO modes in CoSb$_3$ and TlFeCo$_3$Sb$_{12}$.
The values in the unfilled compound are compared to the first-principles
and force-constant (FC) model calculations 
of Feldman and Singh [model A of Ref. \onlinecite{prb53_6273}]
and to the experimental data obtained by infrared~\cite{pssb112_549}
and Raman~\cite{prb59_6189} spectroscopy. 
The overall agreement between theory and experiment is quite good.
Especially the phonon frequencies in the low energy region
--- the region of the Tl vibrations in the filled compound --- are
accurately reproduced by our first-principles calculations.

We should note that the Raman measurements of Ref. \onlinecite{prb59_6189}
were performed on polycrystalline samples. As a consequence, only
the A$_g$ modes were identified in this study while the remaining Raman
lines can be either associated to E$_g$ or F$_g$ modes.
In Table \ref{tab_phon}, we identify four of the additional frequencies
to the theoretical F$_g$ modes. In fact, the experimental frequencies
of 82, 109 and 152 cm$^{-1}$ are close to the three lowest F$_g$ modes
obtained in our study and by Feldman and Singh~\cite{prb53_6273}.
In case of the mode at 178 cm$^{-1}$, this identification is less
clear since its frequency is close to the highest
E$_g$ mode and the highest F$_g$ mode. In addition to
these four modes, the authors of
Ref. \onlinecite{prb59_6189} report a mode at 60 cm$^{-1}$
[not reported in Table \ref{tab_phon}] that is not confirmed by
our calculation neither by the calculation of Feldman
and Singh ~\cite{prb53_6273}. Therefore, we do not associate
this Raman line to a phonon mode in CoSb$_3$.

In spite of the giant effective charges, the transverse
and longitudinal F$_u$ modes are rather close in energy.
This small amplitude of the LO-TO splitting can be attributed
to the large optical dielectric constant 
(see Sec. \ref{sec_eldiel}) that effectively
screens the macroscopic electric field generated by the longitudinal
polar lattice vibrations. At lower energies, this screening is almost
perfect and the LO and TO modes are nearly degenerated. At higher
energies, we observe a small increase in the frequency of the
LO modes with respect to the TO modes. We will see in the
next section that the low and high energy regions are respectively
dominated by vibrations of Sb and Co. The larger amplitude of the
LO-TO splitting at higher energies
can therefore be attributed to the giant effective
charges of Co that are about three times larger than
those of Sb.

In the filled compound TlFeCo$_3$Sb$_{12}$, the presence of
Fe breaks the symmetry of the crystal lattice.
As a consequence, it is no more possible to classify the phonon
modes according to Eq. (\ref{eq_phon}).
Nevertheless, the eigenvectors of most phonons are only slightly
modified in the filled compound.
By computing their overlap,
it is therefore possible to associate each phonon in 
TlFeCo$_3$Sb$_{12}$ to a mode in the unfilled compound.
Due to the symmetry breaking, the degeneracy of some phonons is left
in TlFeCo$_3$Sb$_{12}$. In this case, we report the 
frequencies of all the resulting modes if they differ
by more than one cm$^{-1}$. The first and second values in the last
column of Table \ref{tab_phon} are respectively one and two times
degenerated.

The TO modes that have the strongest overlap with the $F_u^T$ modes
of the unfilled compound, are polar. The single and double degenerate
modes are respectively polarized in the [1 1 1] direction
and in the plane perpendicular to [1 1 1].
The frequencies of the corresponding LO modes are 
reported at the end of Tab. \ref{tab_phon}. In addition to
the $F_u$ modes other modes become polar in the filled
compound. Among them, the 
highest $A_u$ and $E_u$ modes acquire the strongest polarity. Nevertheless,
the splitting between the corresponding TO and LO modes is rather
weak and not reported in Table \ref{tab_phon}.

At 40 cm$^{-1}$, we observe three additional polar modes that are absent
in the unfilled compound. Experimentally, a peak in the phonon
density of states of TlFeCo$_3$Sb$_{12}$ has been observed at this
frequency~\cite{prl90_135505}.
This peak has been attributed to a locallized vibration of Tl in his cage.
Our calculation confirms the strong contribution of Tl to these modes.
About 88 \% of the eigenvectors are due
to the vibrations of Tl. The frequencies of these modes are 
significantly smaller
than the bare frequency of a pure (100 \%) Tl vibration reported in Sec.
\ref{sec_potential} (52 cm$^{-1}$). This difference between the bare
frequency and the frequency of the local Tl mode (LM)
can be attributed to hybridizations between the pure Tl vibrations
and the phonons of the unfilled compound.

To the authors knowledge, no experimental data on the zone-center
phonons of TlFeCo$_3$Sb$_{12}$ are available. However, in a recent
work, Feldman and co-workers~\cite{prb68_94301} measured the frequencies
of the Raman active modes in La$_{0.75}$Fe$_3$CoSb$_{12}$. In addition,
they used first-principles calculations to compute the frequencies of
all zone-center optical phonons in LaFe$_4$Sb$_{12}$. The results of
teir studies are reported in Table \ref{tab_phon} where they are
compared to our theoretical results for TlFeCo$_3$Sb$_{12}$ although
they have been obtained for a different compound. However, the Raman
active modes are dominated by the dynamics of Sb. In 
Sec. \ref{sec_phon_disp}, we will see that the vibrations of Sb are 
almost decoupled of Co/Fe. We can therefore expect the frequencies of
the Raman active modes to be similar in TlFeCo$_3$Sb$_{12}$ and
the La-filled skutterudites. In Table \ref{tab_phon}, we see
that this is indeed the case and that the agreement between
our results and those of Feldman and co-workers is quite good for
most of the Sb-dominated modes between 80 and 200 cm$^{-1}$.
The frequency of the La-dominated mode in LaFe$_4$Sb$_{12}$
is larger than that of the Tl-dominated mode in TlFeCo$_3$Sb$_{12}$.
This result will be discussed more in detail in Sec. \ref{sec_phon_disp}.

\begin{table}[htbp]
\caption{\label{tab_phon} Frequencies (cm$^{-1}$) of the transverse 
and longitudinal zone-center optical phonons in
CoSb$_3$ and TlFeCo$_3$Sb$_{12}$. The phonons in the unfilled compound
are classified according to the irreductible representations of the
space groupe $Im\overline{3}$. Each phonon in TlFeCo$_3$Sb$_{12}$ 
is associated to the mode of the unfilled compound 
for which the overlap between eigenvectors
is the strongest. In case the degeneracy of a mode is left
in the filled compound, we report the frequencies of all resulting 
phonons. The first and second values in column 6 are
respectively one and two times degenerated. The values 
of Ref. \onlinecite{prb68_94301} in the two last
columns have been obtained for La-filled FeSb$_3$ for which the lattice
dynamics is expected to be similar to that in CoSb$_3$.} 
\begin{ruledtabular}
\begin{tabular} {lcrrrr|ccc}
Mode& &
 & \multicolumn{2}{c}{Unfilled} & & 
\multicolumn{3}{c}{Filled} \\
    & & LDA\footnotemark[1] & LDA\footnotemark[2] & FC\footnotemark[2] & Exp. & 
        LDA\footnotemark[1] & LDA\footnotemark[6] & Exp. \\
\hline
LM$^T$& &     &   &   &                   & 41, 42 & 54 &40\footnotemark[5]\\
LM$^L$& &     &   &   &                   & 45     &   \\
A$_g$ & & 141 &150&151&135\footnotemark[3]& 145    & 148 &147\footnotemark[6]\\
      & & 169 &179&177&186\footnotemark[3]& 155    & 156 &154\footnotemark[6]\\
A$_u$ & & 116 &109&110&                   & 112    &  92 &      \\
      & & 236 &241&240&                   & 224    & 212 &      \\
E$_g$ & & 122 &   &139&                   & 123    & 133 &122\footnotemark[6]\\
      & & 175 &   &183&                   & 157    & 158 &161\footnotemark[6]\\
E$_u$ & & 133 &   &131&                   & 130    & 125 &      \\
      & & 246 &   &267&                   & 228    & 251 &      \\
F$_g$ & &  87 &   & 83& 82\footnotemark[3]&  93, 92&  95 & 94\footnotemark[6]\\
      & & 105 &   & 97&109\footnotemark[3]&  98,101& 101 &102\footnotemark[6]\\
      & & 143 &   &157&152\footnotemark[3]& 133,136& 137 &131\footnotemark[6]\\
      & & 173 &   &178&178\footnotemark[3]& 166,163& 164 &172\footnotemark[6]\\
F$^T_u$&&  80 &   & 78& 78\footnotemark[4]&  85    &  94 &      \\
      & & 117 &   &120&120\footnotemark[4]& 116    & 119 &      \\
      & & 136 &   &144&144\footnotemark[4]& 137,135& 140 &      \\
      & & 167 &   &175&174\footnotemark[4]& 152    & 151 &      \\
      & & 232 &   &241&247\footnotemark[4]& 217,215& 223 &      \\
      & & 245 &   &261&257\footnotemark[4]& 248,233& 241 &      \\
      & & 258 &   &277&275\footnotemark[4]& 238,261& 259 &      \\
F$^L_u$&&  80 &   &   & 81\footnotemark[4]&  85    &     &      \\
      & & 119 &   &   &124\footnotemark[4]& 118    &     &      \\
      & & 137 &   &   &147\footnotemark[4]& 137,135&     &      \\
      & & 168 &   &   &                   & 152    &     &      \\
      & & 244 &   &   &252\footnotemark[4]& 233,230&     &      \\
      & & 250 &   &   &262\footnotemark[4]& 244,241&     &      \\
      & & 273 &   &   &288\footnotemark[4]& 267,274&     &      \\
\end{tabular}
\end{ruledtabular}
\footnotetext[1]{Present: CoSb$_3$ and TlFeCo$_3$Sb$_{12}$}
\footnotetext[2]{LDA and force-constnat (FC) model 
                     calculations~\cite{prb53_6273} for CoSb$_3$}
\footnotetext[3]{Raman spectroscopy~\cite{prb59_6189} for CoSb$_3$}
\footnotetext[4]{Infrared spectroscopy~\cite{pssb112_549} for CoSb$_3$}
\footnotetext[5]{Inelastic neutron scattering~\cite{prl90_135505}
                 for Tl-filled CoSb$_3$}
\footnotetext[6]{LDA + Raman data on La-filled 
                 FeSb$_{3}$~\cite{prb68_94301}}
\end{table}

\subsection{Infrared reflectivity and static dielectric tensor}

The static dielectric constant $\varepsilon_0$ 
can be decomposed into an electronic and an ionic
part~\cite{xgcl}
\begin{equation} \label{eq_eps0}
\varepsilon_0 = \varepsilon_{\infty} +
\sum_{m} \Delta \varepsilon_m.
\end{equation}
$\Delta \varepsilon_m$ represents the contribution of an individual
zone-center TO mode and can be computed from the infrared oscillator strenght
$S_m$, the phonon frequency $\omega_m$ and the unit cell volume
$\Omega_0$
\begin{equation} \label{eq_epsm}
\Delta \varepsilon_m = 
\frac{4 \pi}{\Omega_0} \frac{S_m}{\omega_m^2}.
\end{equation}

In Table \ref{tab_oscill}, we report the decomposition of
$\varepsilon_0$ in CoSb$_3$ and TlFeCo$_3$Sb$_{12}$.
In the unfilled compound, the ionic contribution is dominated
by the F$_u$ mode at 232 cm$^{-1}$ that has the strongest oscillator
strength ($S_m = 14.29 \cdot 10^{-4} a. u.$). Such a large
value can be compared to
the oscillator strenghts of the high polar modes in ferroelectric
oxides such as LiNbO$_3$~\cite{prb65_214302}.
In addition, the frequency of this mode is of the
same magnitude as the frequency of the most polar modes in LiNbO$_3$.
However, the 
$\Delta \varepsilon_m$ is small compared to the ionic contribution
to $\varepsilon_0$ in the ferroelectrics. This surprising result
is due to the fact that CoSb$_3$ has a relatively open structure
characterized by a large unit cell volume. In CoSb$_3$, 
the primitive unit cell contains 16 atoms 
and has a volume $\Omega_0 = 369.20 \AA^3$
that is more than three times larger than the volume of the 10 atom
unit cell in LiNbO$_3$ ($\Omega_0 = 101.70 \AA^3$).

\begin{table}[htbp]
\caption{\label{tab_oscill} Electronic and ionic contributions
of individual phonon modes to the static dielectric constant
in CoSb$_3$ and TlFeCo$_3$Sb$_{12}$.}
\begin{ruledtabular}
\begin{tabular} {lrrcrr}
 & \multicolumn{2}{c}{CoSb$_3$} & &
\multicolumn{2}{c}{TlFeCo$_3$Sb$_{12}$} \\
      & $\omega_m$ (cm$^{-1}$) & $\Delta \varepsilon_m$  & &
$\omega_m$ (cm$^{-1}$) &  $\Delta \varepsilon_m$ \\
\hline
Electronic &     & 31.67 & &         & 38.57 \\
\hline
LM         &     &       & &  41, 42 & 10.77 \\
F$_u$      &  80 &  0.09 & &  85     &  0.07 \\
A$_u$      &     &       & & 112     &  0.00 \\
F$_u$      & 117 &  1.76 & & 116     &  2.00 \\
E$_u$      &     &       & & 130     &  0.02 \\
F$_u$      & 136 &  0.38 & & 137,135 &  0.23 \\
F$_u$      & 167 &  0.09 & & 152     &  0.13 \\
F$_u$      & 232 &  6.44 & & 217,215 &  8.04 \\
A$_u$      &     &       & & 224     &  0.93 \\
E$_u$      &     &       & & 228     &  0.06 \\
F$_u$      & 245 &  0.49 & & 248,233 &  0.94 \\
F$_u$      & 258 &  1.52 & & 238,261 &  2.07 \\
\hline
Total      &     & 42.44 & &         & 63.83 \\
\end{tabular}
\end{ruledtabular}
\end{table}

In TlFeCo$_3$Sb$_{12}$, the static dielectric tensor is 
no more isotropic but presents weak off-diagonal elements.
As for the optical dielectric tensor, these elements
are about two orders of magnitude smaller than the diagonal ones.
To a good approximation, we can therefore neglect them
and characterize the dielectric tensor by its single diagonal element
$\varepsilon_0$.
The decomposition of $\varepsilon_0$ in the filled compound
is more complicated than in the unfilled.
The values reported for the F$_u$ modes in Tab. \ref{tab_oscill} 
correspond to the sum of
the $\Delta \varepsilon_m$ of the three modes 
that have been splitted in the filled compound
(see Sec. \ref{sec_phon_freq}).

In TlFeCo$_3$Sb$_{12}$, 
$\varepsilon_0$ is about 50~\% larger than in unfilled CoSb$_3$.
Part of this evolution can be attributed to the electronic
contribution that increases by 6.89 as discussed in Sec.
\ref{sec_eldiel} and to the ionic contribution of the F$_u$
modes that is slightly larger in the filled compound. 
In addition, due to its lower symmetry, the A$_u$ and F$_u$ modes
become infrared active in the filled compound although theyr 
contribution to $\varepsilon_0$ is quite weak.
The most important part (10.77) comes from the Tl
dominated mode at 42 cm$^{-1}$. It's oscillator strength 
($S_m = 0.79 \cdot 10^{-4} a. u.$) is rather low compared to that of the
Co/Fe dominated modes at high energy. 
But due to its low frequency, the
constribution of this additional mode in the filled compound is very
important.
This result demonstrates that Tl-filling has an important 
influence on the static dielectric constant of CoSb$_3$.
It suggests that materials with a structure characterized by
large empty voids might be interesting candidates for applications
that require a well defined dielectric constant since the value
of $\varepsilon_0$ can be tuned by filling the voids (or a fraction of
voids) with foreign atoms.

From the analysis of the results presented in this paper, we can
identify some general criteria that should be satisfied by an element in order
to be efficient for such a tuning of the dielectric constant.
First, the Born effective charge of the
filling atom should be as high as possible. This requires a certain
amount of hybridizations between the foreign atoms and the atoms
of the host crystal. Second, the bare frequency of the 
filling atom should be as low as possible since we expect the normal
mode frequency to increase with the bare frequency.
On the one hand, this criterion requires that, in spite of the 
hybridizations necessary for a high polarizability and Born effective
charge, the filling atom should only be weakly bound
to ensure a low curvature of the potential 
energy well.
On the other hand, the mass of the filling
atom $M$ should be high since the bare frequency is proportional
to $1/ \sqrt{M}$.

\begin{figure}[htb]
\begin{center}
\includegraphics[width=7cm]{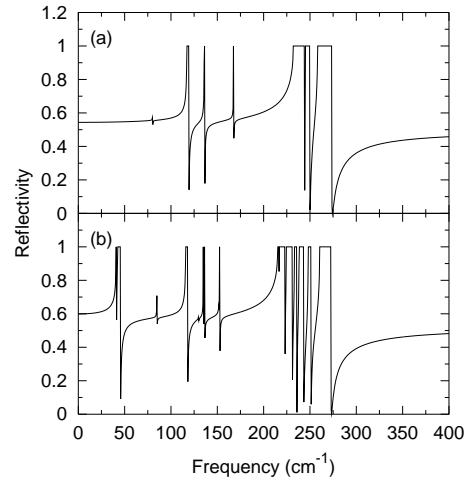}
\end{center}
\caption{\label{fig_refl} Infrared reflectivity of CoSb$_3$ (a) and
TlFeCo$_3$Sb$_{12}$ (b).}
\end{figure}

Fig. \ref{fig_refl} shows the infrared reflectivity of CoSb$_3$ (a)
and TlFeCo$_3$Sb$_{12}$ (b) computed from the Born effective charges
and phonon frequencies and eigenvectors as it is described in Ref.
\onlinecite{xgcl}. Since this approach neglects the damping of the
phonon modes, the curves saturate to 1.
For the unfilled compound, there is a qualitative good agreement
between Fig. \ref{fig_refl} (a) and the reflectivity measured
by Lutz and Kliche~\cite{pssb112_549,irp24_171}.
The most striking difference between the reflectivity of CoSb$_3$
and TlFeCo$_3$Sb$_{12}$ is the presence of a well defined band
below 50 cm$^{-1}$ that is related to the Tl-dominated mode and
the large Tl Born effective charge (Sec. \ref{sec_zeffective}). 
In addition, we observe the appearance of a few lines associated to
the A$_u$ and E$_u$ modes that become infrared active in the filled
compound as discussed above.
Unfortunately,
we are not aware of any experimental reflectivity measurements 
for TlFeCo$_3$Sb$_{12}$. However, Dordevic and co-workers~\cite{prb60_11321}
measured the reflectivity of MFe$_4$P$_{12}$ filled skutterudites
(M = La, Th, Ce, U). They observed a similar strong band associated
to the M-dominated mode in these compounds.
This shows that
infrared spectroscopy is an appropriate
technique to study the dynamics of the filling atoms in skutterudites.

\section{Phonon dispersion curves, density of states and
         specific heat} \label{sec_phon_disp}

\begin{figure*}[htb]
\begin{center}
\includegraphics[width=18cm]{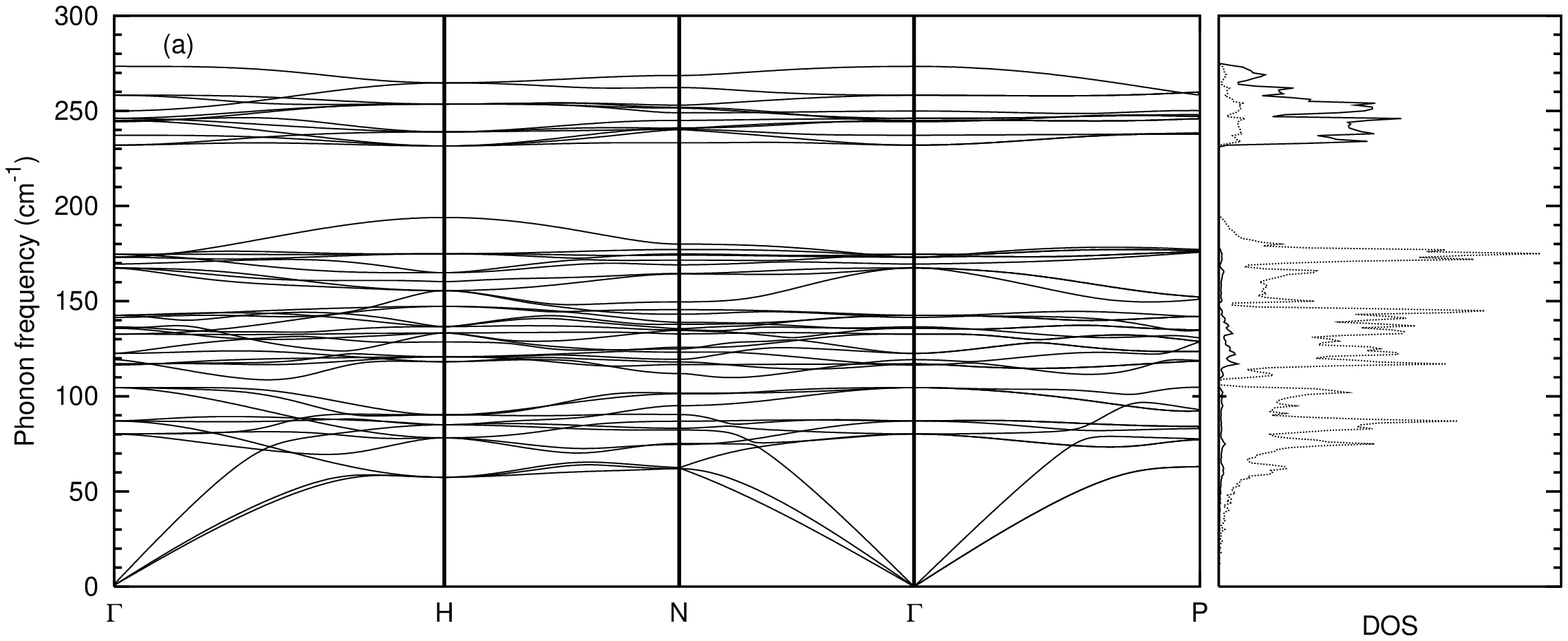}
\includegraphics[width=18cm]{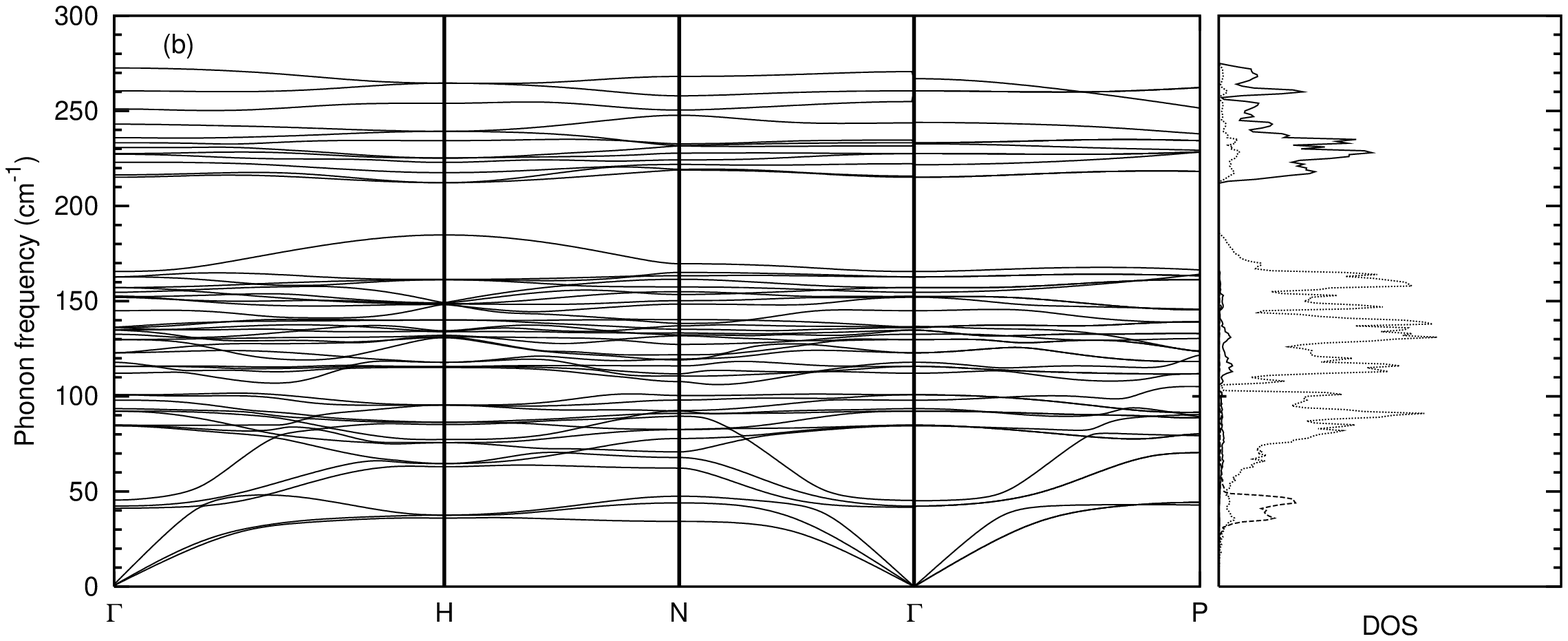}
\end{center}
\caption{\label{fig_phonon_disp}
Phonon dispersion and projected DOS in CoSb$_3$ (a) and 
TlFeCo$_3$Sb$_{12}$ (b). The full, dashed and dotted lines in
the DOS show respectively Co/Fe, Tl and Sb vibrations.}
\end{figure*}

Fig. \ref{fig_phonon_disp} shows the phonon dispersion relation
and projected DOS of CoSb$_3$ (a)
and TlFeCo$_3$Sb$_{12}$ (b). 
In both compounds, the vibrations of Co/Fe and Sb are mainly
decoupled and occupy respectively the high energy region above
220 cm$^{-1}$ and the low energy region below 200 cm$^{-1}$.

The computed DOS of CoSb$_3$ is in good agreement with
the DOS obtained from various force constant model 
calculations by Feldman and Singh~\cite{prb53_6273}.
At low frequency, the agreement is nearly perfect.
All calculations also predict a large
gap at about 200 cm$^{-1}$ and a smaller gap at about 100 cm$^{-1}$.
We do not properly reproduce the two well defined peaks above
200 cm$^{-1}$ reported in Ref.~\onlinecite{prb53_6273} (calculations)
and Ref.~\onlinecite{privatecomm} (experiment).
This might be due to a lack of accuracy in
our calculation of the high energy part of the spectrum.
In Tab. \ref{tab_phon}, we see that the frequencies above
200 cm$^{-1}$ tend to underestimate the experiment while those
below 200 cm$^{-1}$ are more accurate. 
The small deviations at high energy do not affect the following
discussions that focus on the lower part of the spectrum.

In the filled compound, an additional branch at about 40 cm$^{-1}$
gives rise to a well defined peak in the DOS that is not present in
the unfilled compound. A peak at the same frequency has been observed
by inelastic neutron scattering~\cite{prl90_135505}
in the DOS of Tl-filled CoSb$_3$.
By comparing the DOS of the filled and unfilled compound, the
authors of Ref. \onlinecite{prl90_135505}
attributed this peak to a locallized
oscillation of Tl. As can be seen in Fig. \ref{fig_phonon_disp},
the peak at 40 cm$^{-1}$ is strongly dominated by the vibrations
of Tl that show only weak hybridizations with Sb. This result
gives therefore a strong argument in favour of the
interpretation of Ref. \onlinecite{prl90_135505}.

\begin{figure}[htb]
\begin{center}
\includegraphics[width=8cm]{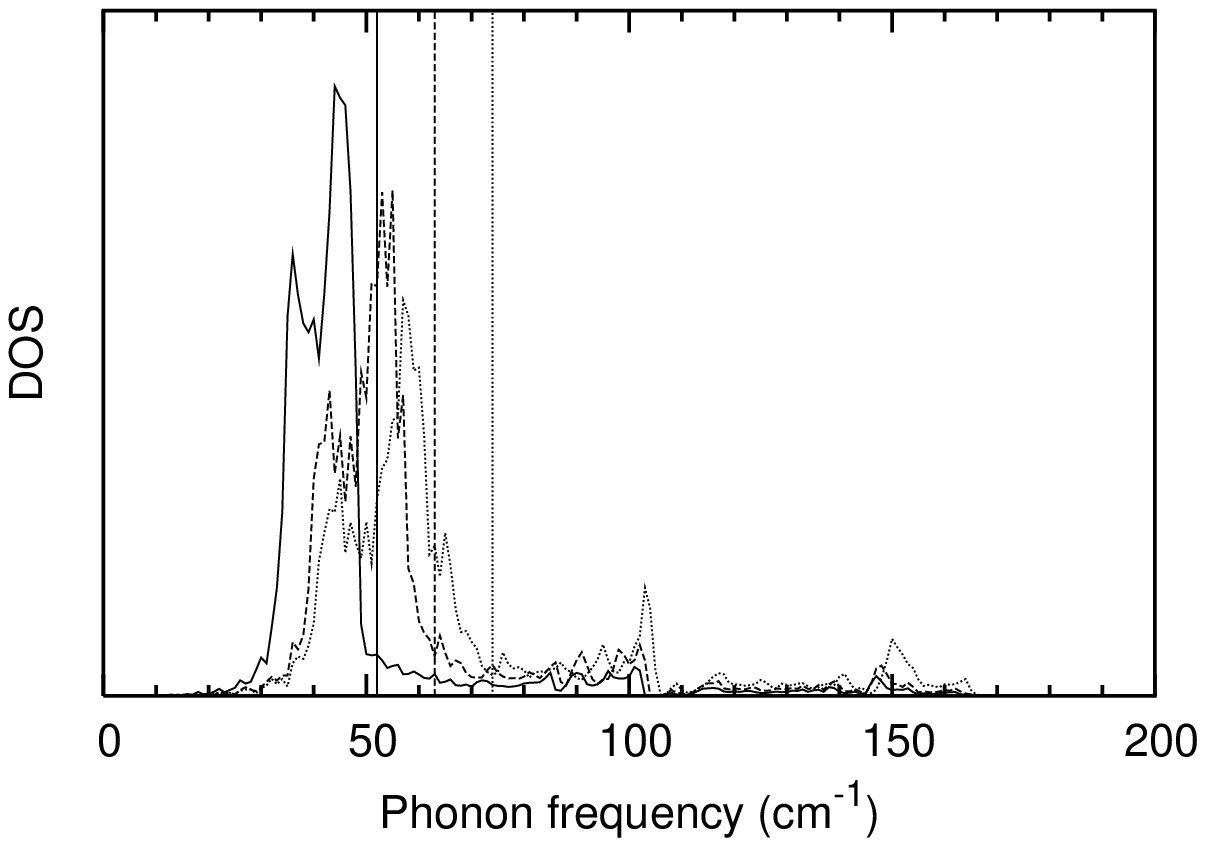}
\end{center}
\caption{\label{fig_dos_tl}
Tl projected DOS in TlFeCo$_3$Sb$_{12}$ (full line), Tl projected
DOS where we artificially decreased the mass of Tl to that of La 
(dashed line) and Tl projected DOS where we changed both the mass
and the curvature of the potential energy well to those of La 
(dotted line). The vertical lines indicate the position of the corresponding
bare frequencies.}
\end{figure}

Experimentally, a difference has been observed between the DOS 
of Tl- and La-filled skutterudites. The authors of Ref. 
\onlinecite{prl90_135505} report the existence of a single Tl-dominated
peak in TlFeCo$_3$Sb$_{12}$ at 40 cm$^{-1}$ while the authos of Ref. 
\onlinecite{nature395_876} claim that there are two La-dominated
peaks in the DOS of LaFe$_4$Sb$_{12}$ at 56 and 121 cm$^{-1}$.
Our calculation clearly confirms the existence of a single peak
in TlFeCo$_3$Sb$_{12}$ (solid line in Fig. \ref{fig_dos_tl}).
In Sec. \ref{sec_potential}, we saw that the bare frequency
of Tl (52 cm$^{-1}$) is significantly smaller than that of La (74 cm$^{-1}$),
a difference that must be attributed to the larger Tl mass
and a weaker curvature of the Tl potential energy well.
If we redo the calculations for a fictitious compound in which we 
artificially reduce the mass of Tl to that of La, the
Tl vibrational frequency is shifted to 48 cm$^{-1}$
and yields a marginal increase in the DOS around 100 cm$^{-1}$
(dashed line in Fig. \ref{fig_dos_tl}).
Further, if we artificially strengthen the interaction of Tl with the lattice by
increasing the interatomic force constants in order to fit the curvature
of the potential energy well of La, the main Tl vibrational peak is shifted
at 51 cm$^{-1}$ whereas we observe the appearance of a
second peak around 100 cm$^{-1}$. This strongly suggests
that the peak at 121 cm$^{-1}$ reported by Keppens and 
co-workers~\cite{nature395_876} cannot be attributed to an additional
localized mode but rather to phonon modes that are formed by coupled
Sb-La vibrations. Generally, it appears that the coupling between
the La and Sb vibrations is more important than between those of Tl and Sb
because in the former case, the bare frequency is closer to the energy region
of the Sb-dominated optical phonons. This interpretation of the second
peak is consistent with the one drawn by Feldman and 
co-workers in Ref. \onlinecite{prb61_r9209}.

As discussed above, the vibrations of Co/Fe are located in the high energy
region of the phonon spectrum so that they are mainly decoupled
from the vibrations of Tl. This result allows us to
justify the approximation used in our study to 
negelect any disorder due to the random substitution of one Co atom per 
cell by Fe (see Sec. \ref{sec_tecdet}). 
Since the vibrations of Tl and Co/Fe are only weakly
coupled, we expect the influence of the disorder on the lattice dynamics
of Tl to be small.
This hypothesis is further confirmed by the observation that the
Tl-peak in the experimental DOS~\cite{prl90_135505} 
is broader in the Sn compensated
compound TlCo$_4$Sb$_{11}$Sn than in TlFeCo$_3$Sb$_{12}$.
In fact, the vibrations of Tl are located in the same energy
region as that of Sb/Sn. As a consequence, there are significant
hybridizations between the vibrations of those atoms.
We expect therefore that the disorder due to the random substitution of one
Sb per cell by Sn has a stronger influence
on the dynamics of the filling atom than that of Co by Fe.

Due to the rather flat dispersion of the optical branches, we expect the
lattice termal conductivity of CoSb$_3$ to be dominated by the acoustic
phonons. As can be seen in Fig. \ref{fig_phonon_disp} (b), the
Tl-derived modes cut through the acoustical branches. A similar effect
has been observed recently in Sr-filled Ge clathrates
from atomic simulations making use of empirical interatomic
potentials~\cite{prl86_2361}. Using molecular dynamics simulations, the authors
of Ref. \onlinecite{prl86_2361} showed that 
the Sr derived modes reduce the lattice thermal conductivity
of the clathrates
by approximately one order of magnitude by scattering the heat-carrying
acoustical phonons. This scattering is enhanced at the frequency
of the Sr-modes because of the resonant interaction with the
acoustical phonons~\cite{prl8_481,prb61_3845,prl82_779}.
It is therefore reasonable to assume that the
resonant interaction between the Tl-derived modes and the
acoustical branches has a
similar strong effect on the thermal conductivity of
CoSb$_3$ and that it is one major source of the important
decrease observed experimentally~\cite{prb61_2475}.

\begin{figure}[htb]
\begin{center}
\includegraphics[width=7cm]{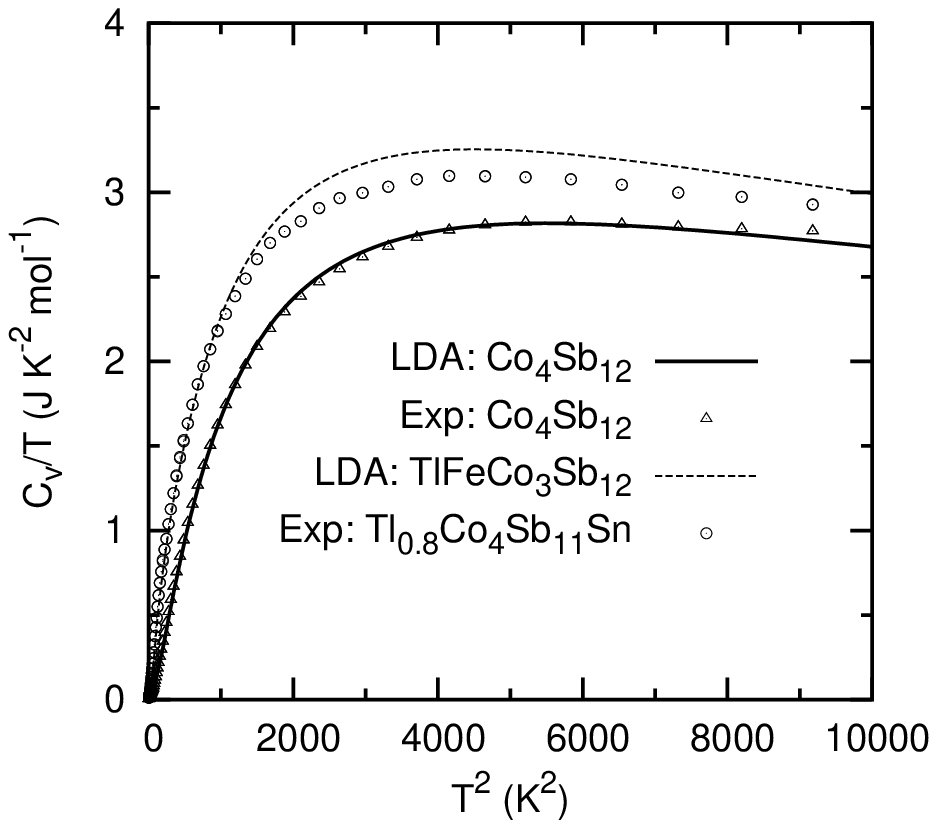}
\end{center}
\caption{\label{fig_heatcapa} Theoretical and experimental
heat capacity of CoSb$_3$ and Tl-filled CoSb$_3$ divided by the temperature
as a function of the square of the temperature. 
Experimental results are from Ref. \onlinecite{prl90_135505}.}
\end{figure}

Fig. \ref{fig_heatcapa} shows the heat capacity at constant volume
of CoSb$_3$ and TlFeCo$_3$Sb$_{12}$ computed by integrating the phonon
DOS as described in Ref. \onlinecite{prb51_8610}.
Our results are compared to the experimental values of 
Ref. \onlinecite{prl90_135505} measured for CoSb$_3$
and Tl$_{0.8}$Co$_4$Sb$_{11}$Sn at constant pressure.
For the unfilled compound, the agreement between theory and experiment
is excellent. In the filled compound, 
we observe a slight deviation from the experimental data.
Part of the discrepancy can probably be attributed to the fact that
the experiment has been performed on a Sn-compensated sample
with only partial void-filling. Especially the substitution of Sb
by Sn is likely to have a stronger effect on the specific heat
at low temperatures than the substitution of Co by Fe since the
Sb/Sn vibrations are at lower energy than that of Co/Fe.

\section{Conclusions}

In this work, we used first-principles DFT calculations to study
the electronic, dielectric and dynamical properties of CoSb$_3$
and TlFeCo$_3$Sb$_{12}$.

Because of hybridizations between Tl and Sb, the Tl 6s electrons
are partially delocalized on the Sb atoms. This result is consistent
with the fact that Tl can be inserted into the voids of CoSb$_3$
inspite of its rather large ionic radius.

The electronic dielectric constant of CoSb$_3$ and TlFeCo$_3$Sb$_{12}$
are quite large with a significantly larger value 
in the filled compound. This increase cannot be
explained in the framework of a Clausius-Mossotti model where the
Tl$^+$ polarizability is supposed to add to the dielectric constant
of CoSb$_3$. Instead, it is the dielectric constant
of the host crystal itself that is increased upon Tl-filling accompagnied
by a substitution of one Co atom per cell by Fe.

Due to the negligible ionic character of the chemical bonds in 
CoSb$_3$ and TlFeCo$_3$Sb$_{12}$,
the {\it static} atomic charges of both compounds
are rather small. In contrast, the {\it dynamic} Born effective charges
are found to be very large and of the same amplitude as in ferroelectric
oxides.

The computation of the energy as a function of a cooperative
displacement of Tl revealed that the center of the cage is the global
minimum of the potential energy and not a saddle point as it
has been suggested previously. In the filled compound, the presence of Tl
gives rise to a {\it single} peak in the phonon DOS. Due
to the hybridizations between the vibrations of Tl and Sb,
this peak is significantly lower in frequency than the Tl bare frequency
computed from the curvature of the potential energy well. However,
these hybridizations are modest compared to those of the La-filled
skutterudites where they give rise to a second La peak in the DOS.

The Tl dominated mode at the $\Gamma$-point is polar and gives rise
to a well defined band in the infrared reflectivity spectrum at low
frequency and a strong increase of the static dielectric constant
in the filled compound. On the one hand, this result shows that infrared
spectroscopy can act as a sensitive probe to study the lattice dynamics
of the filling atom in skutterudites and related compounds such as
clathrates. On the other hand, it suggests that the dielectric constant
of materials characterized by an open structure containing empty voids
can be tuned by filling the voids with foreign atoms.


\section{Acknowledgments}
The authors are grateful to R. P. Hermann, F. Grandjean, G. J. Long
and J. P. Issi
for helpful discussions. M. V. acknowledges
financial support from the FNRS Belgium. This work was supported by
the Volkswagen-Stiftung within the project ``Nano-sized ferroelectric
Hybrids" (I/77 737), the Region Wallonne (Nomade, project 115012),
FNRS-Belgium through grants 9.4539.00 and 2.4562.03, and
the European Commission through the 
FAME Network of Excellence (Functional Advanced Materials Engineering
of Hybrids and Ceramics).



\begin{thebibliography}{}
\bibitem{notediel}
In Ref.~\onlinecite{pssb112_549},
Lutz and Kliche report a value of 3.93 for $|Z^{\ast}_{Co}|$.
However, they used a value of 10.84
for the optical dielectric constant to deduce $|Z^{\ast}_{Co}|$
from the LO-TO splitting of the infrared active phonons. This value
has been corrected to 25.6 in a later 
work~\cite{irp24_171}.
\end{thebibliography}
\end{document}